\documentstyle[11pt]{article}
\begin{document}
\thispagestyle{empty}
\begin{center}
\vspace{1.8cm} {\bf REPRESENTATIONS AND PROPERTIES OF GENERALIZED
$A_r$ STATISTICS}\\ \vspace{1.5cm} {\bf M.
Daoud}\footnote{m$_{-}$daoud@hotmail.com}\\ \vspace{0.5cm} {\it
Facult\'e des Sciences, D\'epartement de Physique, LPMC,\\
Agadir, Morocco}\\[1em]
\vspace{3cm} {\bf Abstract}
\end{center}
\baselineskip=18pt
\medskip
A generalization of $A_r$ statistics is proposed and developed.
The generalized $A_r$ quantum statistics is completely specified
by a set of Jacobson generators satisfying a set of triple
algebraic relations. Fock-Hilbert representations and
Bargmann-Fock realizations are derived.

\newpage
\section{Introduction}
Various kinds of quantum statistics, different from Bose and Fermi
statistics, have been introduced in the literature [1-11]. For
example, in two space dimension, one can have a one parameter
family of statistics (anyons) interpolating between bosons and
fermions [4]. On the other hand, in three and higher space
dimensions the parastatistics, developed by Green [1], constitute
a natural extension of the usual Fermi and Bose statistics. In
this generalization, the classical bilinear commutators or anti-
commutators are replaced by certain triple relations. The trilinear commutation
relations  defining the parafermions are given by
\begin{equation}
\big[[ f_i^+ , f_j^-]  , f_k^- ]\big] = - 2 \delta_{ik}
f_j^- ,{\hskip 0.5cm} \big[[ f_i^+ , f_j^+]  , f_k^- ]\big] = - 2
\delta_{ik}
f_j^+ +  2 \delta_{jk}
f_i^+, {\hskip 0.5cm} \big[[ f_i^- , f_j^-]  , f_k^- ]\big] =
0,
\end{equation}
for $i,j,k = 1,2,...,r$ and the  triple relations
defining the parabosons are defined as
\begin{equation}
\big[\{ b_i^+ , b_j^-\}   , b_k^- ]\big] = - 2 \delta_{ik} b_j^-
,{\hskip 0.5cm}\big[\{ b_i^+ , b_j^+\}   , b_k^- ]\big] = - 2
\delta_{ik} b_j^+ - 2 \delta_{jk}
b_i^+,  {\hskip 0.5cm} \big[\{ b_i^- , b_j^-\}  , b_k^- ]\big]=0.
\end{equation}
It was established that the relations (1)  are associated with the
orthogonal Lie algebra $so(2r+1) = B_r$ [12] and the relations (2)
are connected to the orthosymplectic superalgebra $osp(1/2r) =
B(0,r)$ [13]. Recently, in the spirit of Palev works [6,10], a
classification of generalized quantum statistics were derived for
the classical Lie algebras $A_r$, $B_r$, $C_r$ and $D_r$ [11]. The
increasing interest in generalized statistics stay essentially
motivated by their possible relevance in the understanding of the
theory of fractional quantum Hall effect [14-15], the
superconductivity [16], and probably they can give some adequate
description of black hole statistics [17]. \\ In this paper, We
shall rather focus our attention on a generalization of $A_r$
statistics introduced in [6]. Let us discuss how to realize the
generators and the quantum states, associated with  the
generalized $A_r$ statistics, in a purely analytical way. This
yields, in a natural way, coherent states which are interesting in
many respects [18,19], particularly in the study of dynamical
evolution of quantum systems.\\ The content of this paper is as
follows. Generalized quantum statistics are approached from a set
of Jacobson generators [20] (defined in section 2) satisfying
certain triple relations. This generalization includes two
fundamental sectors. A fermionic one reproducing the $A_r$
statistics introduced in [6]. The second sector approached through
this generalization is of bosonic kind. For each sector, we give
the associated Fock space. An Hamiltonian is derived in terms of
the Jacobson generators identified with creation and annihilation
operators. In section 3, two non equivalents analytic realizations
of the Fock space for the bosonic $A_r$ statistics are performed.
The first realization brings the so-called Gazeau-Klauder coherent
states [21]. The second realization leads to the well known
Klauder-Perelomov coherent states [18,19]. Differential actions of
the Jacobson generators in both realizations are given. In section
4, we realize analytically the Fock space related to the $A_r$
statistics of fermionic kind. In this case, we show that the
Jacobson generators act on a over-complete set of coherent states
similar to Klauder-Perelomov ones labeling the complex projective
spaces ${\bf CP}^r$. Some concluding remarks close this paper.

\section{The generalized $A_r$ statistics}
This section is devoted to the definition of the generalized $A_r$
statistics viewed as Lie triple system. We give the corresponding
Fock space and the Hamiltonian describing a quantum system obeying
the generalized $A_r$ statistics.
\subsection{The algebra}
To begin, let us recall the definition of Lie triple systems. A
vector space with trilinear composition $[x,y,z]$ is called Lie
triple system if the following identities are satisfied:
$$[x,x,x] = 0,$$
$$[x,y,z] + [y,z,x] + [z,x,y] = 0,$$ $$[x,y,[u,v,w]] =
[[x,y,u],v,w] + [u,[x,y,v],w] + [u,v,[x,y,w]].$$\nonumber
Following this definition, we will introduce the generalized $A_r$
statistics as Lie triple system. For this end, we consider the set
of $2r$ operators $x_i^+$ and $x_i^- = (x_i^+)^{\dagger} $, ($i =
1, 2,..., r$) . Inspired by the para-Fermi case [1] and the
example of $A_r$ statistics [6,10], these $2r$ operators should
satisfy certain conditions and relations. First, the operators
$x_i^+$ are mutually commuting. A similar statement holds for the
operators $x_i^-$. They also satisfy  the following triple
relations
\begin{equation}
\big[[ x_i^+ , x_j^- ] , x_k^+ ]\big] = - \epsilon  \delta_{jk}
x_i^+ - \epsilon  \delta_{ij} x_k^+
\end{equation}
\begin{equation}
\big[[ x_i^+ , x_j^- ] , x_k^- ]\big] =  \epsilon  \delta_{ik}
x_j^- + \epsilon  \delta_{ij}  x_k^-
\end{equation}
where $\epsilon \in {\bf R}$. It is easy to see that the algebra
${\cal A}$ (defined by means of the generators $x_i^{\pm}$ and
relations (3) and (4)) is closed under the ternary operation
$[x,y,z] = [[x,y],z]$ and define a Lie triple system. Note that
for $\epsilon = - 1$, the algebra ${\cal A}$ reduces to one
defining the $A_r$ statistics discussed in [6]. It is also
important to remember that the description (3) and (4) of algebras
by means of minimal set of generators and relations was initiated
by N. Jacobson [20]. Therefore, the elements $x_i^{\pm}$ are
referred to as Jacobson generators which will be identified later
with creation and annihilation operators of a quantum system
obeying to generalized $A_r$ statistics. We redefine the
generators of the algebra ${\cal A}$ as $a_i^{\pm} =
\frac{x_i^{\pm}}{\sqrt{|\epsilon|}}$. The triple relations (3) and
(4) rewrite now as
\begin{equation}
\big[[ a_i^+ , a_j^- ] , a_k^+ ]\big] = - s  \delta_{jk}
a_i^+ - s  \delta_{ij} a_k^+
\end{equation}
\begin{equation}
\big[[ a_i^+ , a_j^- ] , a_k^- ]\big] =  s  \delta_{ik}
a_j^- + s \delta_{ij}  a_k^-
\end{equation}
where $ s= \frac{\epsilon}{|\epsilon|}$ is the sign of the
parameter $\epsilon$ and $[ a_i^+ , a_j^+ ] = [ a_i^- , a_j^- ] =
0$. This redefinition is more convenient for our investigation, in
particular in determining the irreducible representation
associated with the algebra ${\cal A}$. The sign of the parameter
$\epsilon$ play an important in the representation of the algebra
${\cal A}$ and consequently, one can obtain different microscopic
and macroscopic statistical properties of the quantum system under
consideration.
\subsection{The Hamiltonian}
To characterize a quantum gas obeying  the generalized $A_r$
statistics, we have to specify the Hamiltonian for the system. The
operators $a_i^{\pm}$ define creation and annihilation operators
for a quantum mechanical system, described by an Hamiltonian $H$,
when the Heisenberg equation of motion
\begin{equation}
[ H , a_i^{\pm} ] = \pm e_i a_i^{\pm}
\end{equation}
is  fulfilled. The quantities $e_i$ are the energies of the modes
$i = 1, 2, ... , r$. One can verify that if $|E \rangle$ is an
eigenstate with energy $E$, $a_i^{\pm}|E \rangle$ are eigenvectors
of $H$ with energies $E \pm e_i$. In this respect, the operators $
a_i^{\pm}$ can be interpreted as ones creating or annihilating
particles. To solve the consistency equation (7), we write the
Hamiltonian $H$ as
\begin{equation}
H = \sum_{i=1}^{r} e_i h_i
\end{equation}
which seems to be a simple sum of "free" (non-interacting)
Hamiltonians $h_i$. However, note that, in the quantum system
under consideration, the statistical interactions occur and are
encoded in the in the triple commutation relations (5) and (6).
Using the structure relations of the algebra ${\cal A}$, the
solution of the Heisenberg condition (7) is given by
\begin{equation}
h_i = \frac{s}{r+1}\bigg[ (r+1)[ a_i^- , a_i^+ ] - \sum_{j=1}^{r}[ a_j^- ,
a_j^+]
\bigg] + c
\end{equation}
where the constant $c$ will be defined later such that the
fundamental state (vacuum) of the Hamiltonian $H$ be zero energy.
\subsection{The representation}
We now consider an Hilbertean representation of the algebra ${\cal
A}$. Let  ${\cal F}$ be the Hilbert-Fock space on which the
generators of ${\cal A}$ act. Since, the algebra ${\cal A}$ is
spanned by $r$ pairs of Jacobson generators, it is natural to
assume that the Fock space is given by
\begin{equation}
{\cal F} = \{ |n_1, n_2,\cdots , n_r\rangle\ , n_i \in {\bf N}\}.
\end{equation}
The action of $a_i^{\pm}$, on ${\cal F}$, are defined as
\begin{equation}
a_i^{\pm} |n_1,\cdots, n_i,\cdots , n_r\rangle\ = \sqrt{F_i
(n_1,\cdots ,n_i \pm 1,\cdots, n_r)}|n_1,\cdots, n_i \pm 1,\cdots
, n_r\rangle\
\end{equation}
in terms of the functions $F_i$ called the structure functions
which should be non-negatives so that all states are well defined.
To determine the expressions of the functions $F_i$ in terms of
the quantum numbers $n_1, n_2, \cdots, n_r$, let first assume the
existence of a vacuum vector $a_i^- |0, 0,\cdots , 0\rangle\ = 0$
for all $i = 1, 2, \cdots, r$. This implies that the functions
$F_i$ can be written
\begin{equation}
F_i (n_1,\cdots ,n_i,\cdots, n_r) = n_i \Delta F_i (n_1,\cdots
,n_i,\cdots, n_r)
\end{equation}
in a factorized form where the new functions $\Delta F_i$ are
defined such that $\Delta F_i (n_1,\cdots ,n_i = 0,\cdots,
n_r)\neq 0$ for $i=1, 2,\cdots, r$. Furthermore, since the
Jacobson operators satisfy the trilinear relations (5) and (6),
these functions should be linear in the quantum numbers $n_i$:
\begin{equation}
\Delta F _i (n_1,\cdots ,n_i,\cdots, n_r)= k_0 + (k_1 n_1 + k_2
n_2 + \cdots + k_r n_r ).
\end{equation}
Finally, using the relations $[a_i^+ , a_j^+] = 0$ and $\big[[
a_i^+ , a_i^- ] , a_i^+ ]\big] = - 2s a_i^+ $, one verifies that $k_i =
k_j $ and $k_i = s$, respectively. For convenience, we set  $k_0 =
k - \frac{1+s}{2}$ assumed to be a non vanishing integer. The
actions of Jacobson generators on the
states spanning the Hilbert-Fock space ${\cal F}$ are now given by
\begin{equation}
a_i^{-} |n_1,\cdots, n_i,\cdots , n_r\rangle\ = \sqrt{n_i (k_0 +
s(n_1+n_2+\cdots+n_r))}|n_1,\cdots, n_i-1,\cdots , n_r\rangle\
\end{equation}
\begin{equation}
a_i^{+} |n_1,\cdots, n_i,\cdots , n_r\rangle\ = \sqrt{(n_i+1) (k_0
+ s(n_1+n_2+\cdots+n_r+1))}|n_1,\cdots, n_i+1,\cdots , n_r\rangle\
\end{equation}
The dimension of the irreducible representation space ${\cal F}$
is determined by the  condition:
\begin{equation}
k_0 + s(n_1+n_2+\cdots+n_r) > 0.
\end{equation}
It depends on the sign of the parameter $s$. It is clear that for
$s=1$, the Fock space ${\cal F}$ is infinite dimension. However,
for $s=-1$, there exists a finite number of states satisfying the
condition $n_1+n_2+\cdots+n_r \leq k-1$. The dimension is given,
in this case, by $\frac{(k-1+r)!}{(k-1)!r!}$. This is exactly the
dimension of the Fock representation of $A_r$ statistics discussed
in [6]. This condition-restriction is closely related to so-called
generalized exclusion Pauli principle according to which no more
than $k-1$ particles can be accommodated in the same quantum
state. In this sense, for $ s = -1$, the generalized $A_r$ quantum
statistics give statistics of fermionic behaviour. They will be
termed here as fermionic $A_r$ statistics and ones corresponding
to $s = 1$ will be named bosonic $A_r$ statistics.\\
Setting $c =
\frac{r}{r+1} s k_0$ in (9) and using the equation (8) together
with the actions of creation and annihilation operators (14-15),
one has
\begin{equation}
H |n_1,\cdots, n_i,\cdots , n_r\rangle\ = \sum_{i=1}^{r}e_i
n_i|n_1,\cdots, n_i,\cdots , n_r\rangle .
\end{equation}
Finally, we point out one interesting property of the generalized
$A_r$ statistics. Introduce the operators $b_i^{\pm} =
\frac{a_i^{\pm}}{\sqrt{k}}$ for $i = 1, 2, \cdots, r$ and consider
$k$ very large. From equations (14) and (15), we obtain
\begin{equation}
b_i^{-} |n_1,\cdots, n_i,\cdots , n_r\rangle\ =
\sqrt{n_i}|n_1,\cdots, n_i-1,\cdots , n_r\rangle\
\end{equation}
\begin{equation}
b_i^{+} |n_1,\cdots, n_i,\cdots , n_r\rangle\ =
\sqrt{n_i+1}|n_1,\cdots, n_i+1,\cdots , n_r\rangle .
\end{equation}
In this limit, the generalized $A_r$ statistics (fermionic and
bosonic ones) coincide with the Bose statistics and the Jacobson
operators reduce to Bose ones (creation and annihilation operators
of harmonic oscillators).\\
Beside the Fock representations
discussed in this section, it is interesting to look for
analytical realizations of the Fock spaces associated with the
generalized $A_r$ statistics. These realizations constitute an
useful tool in connection with variational and path integral
methods to describe the quantum dynamics of the system described
by the Hamiltonian $H$.

\section{Bargmann realizations of  $A_r$ bosonic statistics}
In this section, we give two different realizations \`a la
Bargmann using a suitably defined Hilbert space of entire analytic
functions associated to  $A_r$ statistics of bosonic kind. In the
first analytic realization, the Jacobson creation operators are
realized as simple multiplication by some complex variables. In
the second one, the Jacobson annihilation operators are
represented as derivation with respect to variables of the analytic
representation space. As by product, the first (resp. second)
realization brings, in a natural way, the Gazeau-Klauder
[21](resp. Klauder-Perelomov [18,19]) coherent states associated
to a quantum mechanical system described by the Hamiltonian given
by (8) and obeying the $A_r$ statistics of bosonic kind.
\subsection{ Realization I}
To begin, we ask for a realization
in which the vector $|k;n_1,\cdots,n_r\rangle$ is realized as
powers of complex variables $\omega_1,\cdots,\omega_r$
\begin{equation}
|k; n_1,\cdots,n_r\rangle \longrightarrow C_{k; n_1,\cdots,n_r}
\omega_1^{n_1}\cdots\omega_r^{n_r}
\end{equation}
and Jacobson creation operators $a_i^+$act as a simple
multiplication by $\omega_i$.

The coefficients $C_{k; n_1,\cdots,n_r}$ occurring in (20) are
determined in the following manner. Using  the equation (15) and
the realization (20), one has the recursion relation
\begin{equation}
C_{k; n_1,\cdots,n_i,\cdots,n_r} =
((n_i+1)(k+n_1+\cdots+n_i+\cdots+n_r))^{\frac{1}{2}}
C_{k;n_1,\cdots,n_i+1,\cdots,n_r},
\end{equation}
which can be easily solved
\begin{equation}
C_{k; n_1,\cdots,n_i,\cdots,n_r} = \bigg[{\frac{(k-1 + n - n_i)!
}{n_i!(k-1 + n) !}}\bigg]^{\frac{1}{2}} C_{k;
n_1,\cdots,0,\cdots,n_r}
\end{equation}
where $n = n_1 + n_2 + \cdots + n_r$. Repeating this procedure for
all $i = 1, 2,\cdots,r$ and setting $C_{k; 0,\cdots,0} = 1$, we
obtain
\begin{equation}
C_{k; n_1,\cdots,n_i,\cdots,n_r} = \bigg[{\frac{(k-1)!
}{n_1!\cdots n_r!(k-1 + n) !}}\bigg]^{\frac{1}{2}}.
\end{equation}
The differential realization of the operators $N_i$ ($\neq a_i^+
a_i^-$) acting on the Fock space ${\cal F}$ as
\begin{equation}
N_i|k;n_1,\cdots,n_i\cdots,n_r\rangle = n_i
|k;n_1,\cdots,n_i\cdots,n_r\rangle,
\end{equation}
are given by
\begin{equation}
N_i \longrightarrow \omega_i \frac{d}{d\omega_i}.
\end{equation}
Using the realization (25), the actions of $a_i^-$ on ${\cal F}$
and the relations defining the algebra ${\cal A}$, one show that
the annihilation operators act as
\begin{equation}
a_i^- \longrightarrow k \frac{d}{d\omega_i} + \omega_i
\frac{d^2}{d^2\omega_i} + \frac{d}{d\omega_i}\sum_{i \neq
j}\omega_j \frac{d}{d\omega_j}
\end{equation}
a second order differential on the Bargmann space generated by the
set of complex variables $\omega_1,\cdots,\omega_r$. A general
vector
\begin{equation}
|\psi\rangle = \sum_{n_1,\cdots,n_r}
\psi_{n_1,\cdots,n_r}|k;n_1,\cdots,n_r\rangle
\end{equation}
in the Fock space ${\cal F}$ now is realized as follows
\begin{equation}
\psi(\omega_1,\cdots,\omega_r) = \sum_{n_1,\cdots,n_r}
\psi_{n_1,\cdots,n_r}C_{k;n_1,\cdots,n_r}\omega_1^{n_1}\cdots\omega_r^{n_r}.
\end{equation}
We define the inner product in this realization in the following
form
\begin{equation}
\langle\psi'|\psi\rangle = \int d^2\omega_1 \cdots d^2\omega_r
K(k;\omega_1,\cdots,\omega_r)\psi'^{\star}(\omega_1,\cdots,\omega_r)
\psi(\omega_1,\cdots,\omega_r)
\end{equation}
where $d^2\omega_i \equiv dRe\omega_idIm\omega_i$ and the
integration extends over the entire complex space $\bf{C}^r$. To
compute the measure function, appearing in the definition of the
inner product (29), we identify  $|\psi\rangle$(resp.
$|\psi'\rangle$) with the vector $|k;n_1,\cdots,n_r\rangle$ (resp.
$|k;n'_1,\cdots,n'_r\rangle$). We also assume that it depends only
on $\rho_i = |\omega_i|$ for $i = 1,\cdots, r$. This hypothesis,
usually used in the moment problems, is called the isotropy
condition. It is a simple matter of computation to show that the
function $K(k;\rho_1,\cdots,\rho_r)$ satisfy the integral
equation
\begin{equation}
(2\pi)^r\int_0^{\infty}\cdots\int_0^{\infty} d\rho_1\cdots d\rho_r
K(k;\rho_1,\cdots ,\rho_r) |\rho_1|^{2n_1+1}\cdots
|\rho_r|^{2n_r+1}= \frac{n_1!\cdots n_r!(k-1+n)!}{(k-1)!}.
\end{equation}
A solution of this equation exists [22](see a nice proof in [23])
in term of the Bessel function
\begin{equation}
K(k;R) = \frac{2}{\pi^r(k-1)!}R^{k-r}K_{k-r}(2R)
\end{equation}
where $R^2 = \rho_1^2 + \cdots + \rho_r^2$. Note that the analytic
function $\psi (\omega_1, \cdots ,\omega_r)$ can be viewed as the
inner product of the ket $|\psi\rangle$ with a bra $\langle k;
\omega_1^{\star}, \cdots ,\omega_r^{\star}|$ labeled by the
complex conjugate of the variables $\omega_1, \cdots ,\omega_r$
\begin{equation}
\psi (\omega_1, \cdots ,\omega_r) = {\cal N}\langle k;
\omega_1^{\star}, \cdots ,\omega_r^{\star}|\psi \rangle
\end{equation}
where ${\cal N} \equiv {\cal N}(|\omega_1|, \cdots ,|\omega_r|)$
stands for a normalization constant. As particular case, we take
$|\psi \rangle = |k; n_1,\cdots,n_r\rangle$ to obtain
\begin{equation}
\langle k; \omega_1^{\star}, \cdots ,\omega_r^{\star}|k;
n_1,\cdots,n_r \rangle = {\cal N}^{-1} C_{k;n_1,\cdots,n_r}
\omega_1^{n_1} \cdots \omega_r^{n_r}.
\end{equation}
It follows that the normalized states $|k; \omega_1, \cdots
,\omega_r\rangle$ are expressed as
\begin{equation}
|k; \omega_1, \cdots ,\omega_r\rangle = {\cal N}^{-1}
\sum_{n_1=0}^{\infty} \cdots \sum_{n_r=0}^{\infty}
\bigg[{\frac{(k-1)! }{n_1!\cdots n_r!(k-1 + n)
!}}\bigg]^{\frac{1}{2}} \omega_1^{n_1} \cdots \omega_r^{n_r} |k;
n_1, \cdots ,n_r\rangle
\end{equation}
where
\begin{equation}
{\cal N}^2(|\omega_1|, \cdots ,|\omega_r|) = \sum_{n_1=0}^{\infty}
\cdots \sum_{n_r=0}^{\infty} {\frac{(k-1)! }{n_1!\cdots n_r!(k-1 +
n) !}} |\omega_1|^{2n_1} \cdots |\omega_r|^{2n_r}.
\end{equation}
It may be noted that the states $|k; \omega_1, \cdots
,\omega_r\rangle$ are not orthogonal. It is also interesting to
remark that they are eigenvectors of the Jacobson operators
$a_i^-$ with the eigenvalue $\omega_i$. In this sense, the states
$|k; \omega_1, \cdots ,\omega_r\rangle$ can be considered as
Gazeau-Klauder  coherent states associated with  a quantum
mechanical system whose Hamiltonian is given by (8) and obeying
the $A_r$ bosonic statistics. Finally, note that this first
realization is possible because the Fock space is infinite
dimension.
\subsection{ Realization II}
There exists a second (non equivalent) Bargmann realization
associated with $A_r$ bosonic systems. In this realization the
annihilation operators act as derivation with respect to complex
variables $z_i$
\begin{equation}
a_i^-\longrightarrow \frac{d}{dz_i}
\end{equation}
belonging to the complex domain ${\cal D} = \{(z_1, z_2, \cdots,
z_r ): |z_1|^2 + |z_2|^2 + \cdots + |z_r|^2 < 1 \}$ . The reason
of this condition will explained in the sequel of this section.
The elements of the Fock space are realized as follows
\begin{equation}
|k; n_1, \cdots , n_r\rangle  \longrightarrow C_{k; n_1,\cdots
,n_r} z_1^{n_1} \cdots z_r^{n_r}
\end{equation}
Using the action of the annihilation operators on the Fock space
${\cal F}$ and the correspondence (37), one obtain the following
recursion formula
\begin{equation}
\sqrt{k-1+n_1+\cdots +n_i+ \cdots +n_r}C_{k; n_1,\cdots,n_i-1,
\cdots,n_r}= \sqrt{n_i}C_{k; n_1,\cdots,n_i,
\cdots,n_r}
\end{equation}
which can be solved in a similar manner that one used above (Eq.
21). As result, we obtain
\begin{equation}
C_{k; n_1,\cdots,n_i,
\cdots,n_r}= \bigg[{\frac{(k-1 + n)!
}{n_1!\cdots n_r! (k-1)!}}\bigg]^{\frac{1}{2}}
\end{equation}
where $n = n_1 + n_2 + \cdots + n_r$. Having the expression of the
coefficients $C$, one can determine the differential action of the
Jacobson creation operators. Indeed, using the actions of the
generators $a_i^+$ on the Fock space, we show that
\begin{equation}
a_i^+ \longrightarrow k z_i + z_i \sum_{j=1}^r z_j\frac{d}{dz_j}.
\end{equation}
In this realization, the Jacobson generators act as first order
linear differential operators. Here also, we realize a general
vector of the Fock space ${\cal F}$ $$|\phi \rangle =
\sum_{n_1=0}^{\infty}\sum_{n_2=0}^{\infty}\cdots
\sum_{n_r=0}^{\infty} \phi_{n_1, n_2 \cdots , n_r}|k; n_1,n_2,
\cdots ,n_r \rangle$$\nonumber\\ as
\begin{equation}
\phi(z_1, z_2 \cdots , z_r) =
\sum_{n_1=0}^{\infty}\sum_{n_2=0}^{\infty}\cdots
\sum_{n_r=0}^{\infty} \phi_{n_1, n_2 \cdots , n_r} C_{k;n_1, n_2
\cdots , n_r} z_1^{n_1} z_2^{n_2} \cdots z_r^{n_r}.
\end{equation}
The inner product of two functions $\phi$ and $\phi'$ is defined
now as follows
\begin{equation}
\langle\phi'|\phi\rangle = \int \int \cdots \int d^2z_1  d^2z_2
\cdots d^2z_r \Sigma(k;z_1,z_2 \cdots ,z_r)\phi'^{\star}(z_1,z_2
\cdots ,z_r) \phi(z_1,z_2 \cdots ,z_r)
\end{equation}
where the integration is carried out the complex domain ${\cal
D}$. The computation of the integration measure $\Sigma$, assumed
to be isotropic , can be performed by choosing $|\phi\rangle = |k;
n_1,n_2, \cdots ,n_r \rangle $ and $|\phi'\rangle = |k; n'_1,n'_2,
\cdots ,n'_r \rangle $. So, one has to look for a solution of the
following moment equation
\begin{equation}
\int \int \cdots \int d\varrho_1  d\varrho_2
\cdots d\varrho_r \Sigma(k;\varrho_1,\varrho_2, \cdots ,\varrho_r)
\varrho_1^{2n_1+1}\varrho_2^{2n_2+1} \cdots \varrho_r^{2n_r+1} =
\frac{n_1!n_2!\cdots n_r! (k-1)!}{(2\pi)^r(k-1+n)!}
\end{equation}
where $n = n_1+n_2+\cdots +n_r$ and $\varrho_i = |z_i|$. To find
the function satisfying the equation (43), we use following result
$$\int_0^1 \varrho_1^{n_1}d\varrho_1  \int_0^{1-\varrho_1}
\varrho_2^{n_2}d\varrho_2 \cdots  \int_0^{1-\varrho_1-\varrho_2-\cdots
-\varrho_{r-1}}
\varrho_r^{n_r}(1-\varrho_1-\varrho_2-\cdots
-\varrho_r)^{k-r-1}d\varrho_r$$ \nonumber
\begin{equation}
= \frac{n_1!n_2!\cdots n_r!(k-1)!}{(k-1+n)! (k-r)(k-r+1)\cdots
(k-1)}
\end{equation}
which can be easily verified. The measure is then given by
\begin{equation}
\Sigma (k;\varrho_1,\varrho_2, \cdots ,\varrho_r) = \pi^{-r}[1 -
(\varrho_1^2 + \varrho_2^2+ \cdots +
\varrho_r^2)]^{k-r-1}(k-r)(k-r+1)\cdots (k-1).
\end{equation}
At this stage, one can write the function $\phi(z_1, z_2 \cdots ,
z_r)$ as the product of the state $|\phi\rangle $ with some ket
$|k; z^*_1, z^*_2 \cdots , z^*_r \rangle$ labeled by the complex
conjugate of the variables $z_1, z_2, \cdots , z_r$.
\begin{equation}
\phi(z_1, z_2, \cdots , z_r)= {\cal N} \langle k; z^*_1, z^*_2,
\cdots , z^*_r |\phi \rangle.
\end{equation}
Taking $|\phi\rangle = |k; n_1, n_2, \cdots , n_r \rangle$, we
obtain
\begin{equation}
|k; z_1, z_2, \cdots , z_r \rangle =  {\cal N}^{-1}
\sum_{n_1=0}^{\infty}\sum_{n_2=0}^{\infty}\cdots
\sum_{n_r=0}^{\infty} \bigg[{\frac{(k-1 + n)! }{n_1!\cdots n_r!
(k-1)!}}\bigg]^{\frac{1}{2}} z_1^{n_1} z_2^{n_2} \cdots z_r^{n_r}
|k; n_1, n_2, \cdots , n_r \rangle.
\end{equation}
The expansion in (47) converges when the the complex variables
$z_1, z_2, \cdots , z_r$ are in the $r$-dimensional unit ball
$|z_1|^2+ |z_2|^2+ \cdots + |z_r|^2 < 1$. This explain the above
hypothesis according to which $z_1, z_2, \cdots , z_r$ should be
in the complex domain ${\cal D}$. The normalization condition of
the states (47) gives the following expression for the
normalization constant
\begin{equation}
{\cal N} = (1 - |z_1|^2- |z_2|^2- \cdots - |z_r|^2)^\frac{k}{2}.
\end{equation}
It is important to remark that the states (47) are continuous in
the labeling, constitute an over complete set and then are
coherent in the Klauder-Perelomov sense [18,19]. These states can
be used in the investigation of quantum Hall systems in higher
dimensional ball [24,25].

\section{Bargmann realization of $A_r$ fermionic statistics}
We are going to construct the analytic realization of the
irreducible representation, related to $A_r$ fermionic statistics
$(s = -1)$ characterized by the so-called the generalized Pauli
principle. As we mentioned before, because the Fock space is
finite dimension, the Jacobson creation operators can not be
represented by a multiplication by some complex variable and only
one analytic realization can be established. It corresponds to one
in which the generators $a_i^-$ act as
\begin{equation}
a_i^- \longrightarrow \frac{d}{d\zeta_i}
\end{equation}
in the space of the polynomials of the form $C_{k; n_1, n_2,\cdots
, n_r } \zeta_1^{n_1}\zeta_2^{n_2} \cdots \zeta_r^{n_r}$ defined
in the $r$-dimensional space $\bf{C}^r$of complex lines $(\zeta_1,
\zeta_2, \cdots , \zeta_r)$ with the correspondence
\begin{equation}
|k; n_1, n_2,\cdots , n_r  \rangle\longrightarrow C_{k; n_1,
n_2,\cdots , n_r } \zeta_1^{n_1}\zeta_2^{n_2} \cdots
\zeta_r^{n_r}.
\end{equation}
Following similar analysis as above, one can see that the
coefficients in (50) satisfy the recurrence formula
\begin{equation}
\sqrt{n_i}C_{k; n_1, n_2,\cdots ,n_i,\cdots ,n_r } =\sqrt{k-n}
C_{k; n_1, n_2,\cdots ,n_i-1,\cdots , n_r }
\end{equation}
where $n = n_1 + n_2 + \cdots + n_r$. Solution of this equation,
for all $i=1, 2, \cdots , r$, is given by
\begin{equation}
C_{k; n_1, n_2,\cdots ,n_r } = \bigg[ \frac{(k-1)!}{n_1!n_2!\cdots
n_r!(k-1-r)} \bigg]^{\frac{1}{2}}
\end{equation}
Using the action of the creation operators on the Fock space, it
is not difficult to verify that they act
\begin{equation}
a_i^+ \longrightarrow (k-1)\zeta_i - \zeta_i \sum_{j=1}^{r}\zeta_j
\frac{d}{d\zeta_i}
\end{equation}
as first order differential operators. As in the previous
realizations, the existence of a measure $\sigma (k; \zeta_1,
\zeta_2, \cdots , \zeta_r)$ ensures the definition of inner
product of two functions. To determine this measure, one use the
orthogonality of the Fock states $|k; n_1, n_2, \cdots ,n_r
\rangle$. It follows that it satisfies $$\int \int \cdots \int
d^2\zeta_1 d^2\zeta_2 \cdots d^2\zeta_r \sigma (k; \zeta_1,
\zeta_2, \cdots , \zeta_r) C_{k; n_1, n_2, \cdots ,n_r}C_{k; n'_1,
n'_2, \cdots ,n'_r} \zeta_1^{n_1}\zeta_1^{n'_1}
\zeta_2^{n_2}\zeta_2^{n'_2} \cdots
\zeta_r^{n_r}\zeta_r^{n'_r}$$\nonumber
\begin{equation}
= \delta_{n_1,n'_1}\delta_{n_2,n'_2}
\cdots  \delta_{n_r,n'_r}.
\end{equation}
Setting $\zeta_i = |\zeta_i|e^{i\theta}$ and assuming  the
isotropy condition, the last equation becomes
$$ \int_0^{\infty}
\int_0^{\infty} \cdots \int_0^{\infty}dx_1 dx_2 \cdots dx_r\mu(k,
x_1 ,x_2, \cdots ,x_r) x_1^{n_1}x_2^{n_2}\cdots x_r^{n_r} $$
\begin{equation}
= \frac{n_1!n_2!\cdots n_r!(k-1-n)!}{(k-1)!}
\end{equation}
where $\mu \equiv \pi^r \sigma$ and $x_i = |\zeta_i|^2$. Using the
Mellin inverse transform [22], we obtain
\begin{equation}
\mu(k, x_1 ,x_2, \cdots ,x_r) = \frac{(k-1+r)!}{(k-1)!}(1 + x_1 +
x_2 + \cdots + x_r )^{-(k+r)}.
\end{equation}
Any function $f(\zeta_1, \zeta_2, \cdots , \zeta_r)$ can be
written in the following form
\begin{equation}
f(\zeta_1, \zeta_2, \cdots , \zeta_r) = {\cal N} \langle k;
\zeta_1^*, \zeta_2^*, \cdots , \zeta_r^* | f \rangle
\end{equation}
where $| f \rangle$ is a generic element of the Fock space with $s
= -1 $ and the normalization constant is given by
\begin{equation}
{\cal N}(|\zeta_1|^2 , |\zeta_2|^2 , \cdots , |\zeta_r|^2) = (1 +
|\zeta_1|^2 + |\zeta_2|^2 + \cdots + |\zeta_r|^2
)^{-\frac{k-1}{2}}.
\end{equation}
It is remarkable to notice that the states $ |k; \zeta_1, \zeta_2,
\cdots , \zeta_r\rangle$ are nothing but the coherent states for
the complex projective space $\mathbf{CP}^r$ which emerges, in a
natural way, in realizing analytically the quantum states of a
system obeying fermionic $A_r$ statistics. It is also interesting
to mention that the states $ |k; \zeta_1, \zeta_2, \cdots ,
\zeta_r\rangle$ were used in the description of quantum Hall
systems in higher dimension complex projective spaces [24]. I this
respect, the generalized quantum $A_r$ statistics can be applied
in the description of quantum Hall effect in $\mathbf{CP}^r$
spaces.
\section{Concluding remarks}
The main ingredient for the present paper is the definition of
generalized $A_r$ statistics. We have obtained the Fock
representation associated to bosonic ($s = 1$) and fermionic ($s =
-1 $) $A_r$ statistics. We have developed the Bargmann
realizations of Fock spaces and the differential action of
Jacobson generators. We shown that the analytical representations
lead to over-complete sets identified as Klauder-Perelomov and
Gazeau-Klauder coherent states associated with a quantum system
described by the Hamiltonian (8). The present work can be extended
to quantum statistics associated with classical Lie algebras
$B_r$, $C_r$ and $D_r$. It will also be interesting to study the
macroscopic properties of such statistics. Another possible
extension concerns the connection of the generalized $A_r$
statistics and the quantum Hall effect in higher dimension spaces.

\end{document}